\documentclass[amsmath,aps,prb,showpacs,preprint]{revtex4}
\begin{document}

\title{Transmission, reflection and localization in a random medium with absorption or gain}
\author{Jean Heinrichs}
\email{J.Heinrichs@ulg.ac.be}
\affiliation{Institut de Physique, B{5}, Universit\'{e} de Li\`{e}ge, Sart
Tilman, B-{4000} Li\`{e}ge, Belgium}
\date{\today}

\begin{abstract}
We study reflection and transmission of waves in a random tight-binding system with absorption or gain for weak disorder, 
using a scattering matrix formalism.  Our aim is to discuss analytically the effects of absorption or gain on the
statistics of wave transport.  Treating the effects of absorption or gain exactly in the limit of no disorder, allows us
to identify short- and long lengths regimes relative to absorption- or gain lengths, where the effects of absorption/gain
on statistical properties are essentially different.  In the long-lengths regime we find that a weak absorption or a weak
gain induce identical statistical corrections in the inverse localization length, but lead to different corrections in the
mean reflection coefficient.  In contrast, a strong absorption or a strong gain strongly suppress the effect of disorder
in identical ways (to leading order), both in the localization length and in the mean reflection coefficient.
\end{abstract}


\maketitle

\section{INTRODUCTION}

In this paper we study analytically the coherent reflection and transmission of waves in an active one-dimensional 
disordered system which either absorbs or amplifies the waves.  Our model is the familiar single-band tight-binding model
with random site energies (Anderson model) including additional fixed positive or negative imaginary parts describing
absorption or amplification.  As is well-known, the introduction of the imaginary potential destroys the time-reversal
symmetry of the system.

The electronic model with absorption may describe annihilation of electrons via electron-hole recombinations acting as a 
complex optical potential in a nearly compensated semiconductor.  The amplification model is, of course, meaningless for
electrons whose fermionic character forbids the presence of more than one electron at a given spatial location.

On the other hand, the tight-binding model with absorption or amplification due to stimulated emission may be appropriate 
for describing the localization of light waves in active photonic band-gap crystals, characterized by a periodic variation
of the dielectric constant.  In particular, the interplay of the (phase coherent) amplification of light waves with the
process of coherent scattering by random inhomogeneities leading to localization\cite{1,2} is of current interest for
random lasers\cite{3}.

A considerable amount of theoretical work related to the statistics of the transmittance and of the reflectance of random 
systems with absorption or gain has already been published\cite{4,5,6,7,8,9,10,11,12,13,14,15}.  We feel, however, that
important aspects of the effects of absorption or gain on the transmission properties of the random system studied below,
have not received sufficient attention in previous work.

We refer, in particular, to the form of the localization length $\xi$ for large lengths $L$ of a random chain.  For weak 
disorder this is found to be given by\cite{6,9}

\begin{equation}\label{new1}
\frac{1}{\xi}=\frac{1}{l_u}+\frac{1}{\xi_0},u=a\;\text{or}\; g
\quad ,
\end{equation}
where $\xi_0$ is the localization length of the system in the absence of absorption or gain (amplification), $l_a$ and 
$l_g$ are the absorption and gain lengths of a perfect system, respectively.  In the tight-binding model, (\ref{new1}) is
expected to be obtained by studying the transmittance of the disordered sample described by the Schr\"{o}dinger equation

\begin{equation}\label{new2}
[E-(\varepsilon_n+i\eta)]\varphi_n=\varphi_{n+1}+\varphi_{n-1}, \; n=1,2,\ldots, N
\quad ,
\end{equation}
where $\varphi_m$ are the wavefunction amplitudes at sites $m=1,2,\ldots, N $, of spacing $a$, of the disordered sample 
of length $L=Na$.  The $\varepsilon_m$ are the random site energies to which one adds a fixed non-hermitian term $i\eta$
describing absorption for $\eta>0$ and amplification for $\eta<0$.  The energies $E$, $\varepsilon_m$ and $\eta$ are in
units of a fixed nearest-neighbour hopping energy $V$.  The random chain is connected at both ends to semi-infinite
perfect leads ($\varepsilon_m=0$) with $\eta=0$, whose sites are positioned at 
$m=0,-1,-2,\ldots$ and $m=N+1,N+2,\ldots$, respectively.
For the perfect tight-binding chain with absorption or gain, Datta\cite{12} has derived 

\begin{equation}\label{new3}
l_g=l_a=\frac{1}{|\eta|}
\quad ,
\end{equation}
while numerical studies of the random tight-binding model by Gupta, Joshi and Jayannavar\cite{11} and by Jiang and 
Soukoulis\cite{13} support Eq. (\ref{new1})  for small $\eta$ of either sign, with $1/\xi_0$ given by the familiar
Thouless expression for weak site-energy disorder.  A feature of (\ref{new1}) which is generally regarded as paradoxical
is the fact that it leads to suppression of transmittance for large $L$, as in the case of absorption.  What is more,
according to equation (\ref{new1}), when disorder is present the suppression for amplification occurs at exactly the same
rate as for absorption\cite{9}.  This surprising feature has been an important incentive for developing a more
comprehensive analytic treatment of the statistics of wave transport in the presence of absorption or amplification. 
Indeed, an important drawback of Eq. (\ref{new1}) is that it completely ignores effects of absorption or gain on the
statitstics of the transmission coefficient from which $\frac{1}{\xi}$ is obtained.  Our aim is to remedy to this defect
in the framework of a detailed analysis for weak disorder of the tight-binding system defined by Eq. (\ref{new2}).

An important feature of our approach below is an exact treatment of the effect of absorption or gain, as done previously 
by Datta\cite{12} for a non-disordered (pure) system.  This allows us to clearly identify and discuss short- and long
lengths regimes relative to the absorption (amplification) length in (\ref{new1}).

The study of transmission and reflection in random one-dimensional media with absorption or gain was initiated, and later
 pursued actively, using invariant imbedding equations\cite{4,5,6,7,8,9,10}.  These equations are coupled non-linear
differential equations for the reflection- and transmission amplitudes of plane waves incident at the right of a
continuous medium occupying the domain $0\leq x\leq L$ of the $x$-direction\cite{16}.  As was shown recently\cite{17}, the
invariant imbedding equations follow for weak disorder from the long-wavelength continuum limit of (\ref{new2}) for a
disordered chain embedded in an infinite perfect chain.  We recall that the invariant imbedding equations were originally
derived as an exact consequence of the Helmholtz equation for the propagation of the electric field in a dielectric
medium\cite{4,18,19}.  For later discussion, we also recall the important early result for the so-called short-lengths
localization length\cite{4,5},

\begin{equation}\label{new4}
\frac{1}{\xi}=\eta+\frac{1}{\xi_0}
\quad ,
\end{equation}
which indicates that at short lengths transmittance increases with $L$ for amplification, if $l_a<\xi_0$.

The paper is organized as follows.  In Sect. II we derive exact expressions for the scattering matrix elements 
(transmission and reflection amplitudes) in terms of transfer matrices for the tight-binding equation (\ref{new1}) for
weak disorder.  In Sect. III we discuss our explicit analytic results for the averaged logarithmic transmission
coefficients (localization lengths) in the short- and long lengths regimes.  We also discuss the mean logarithmic
reflection coefficient for large lengths.  We recall that the distribution of the reflection coefficient is important in
the context of random lasers\cite{7,8}.  Some concluding remarks follow in Sect. IV.

\section{TRANSFER MATRIX ANALYSIS}

We start our analysis by rewriting (\ref{new2}) in terms of a transfer matrix for a site $n$,

\begin{equation}\label{new5}
\begin{pmatrix}
\varphi_{n+1}\\
\varphi_n
\end{pmatrix}=
\widehat P_n
\begin{pmatrix}
\varphi_n\\
\varphi_{n-1}
\end{pmatrix},\quad
\widehat P_n=
\begin{pmatrix}
E-\varepsilon_n-i\eta & -1\\1 & 0
\end{pmatrix}
\quad .
\end{equation}
The analogous equation for sites in the perfect leads involves the transfer matrix $\widehat P^0$ obtained by letting 
$\varepsilon_n=\eta=0$ in(\ref{new5}).  We wish to study the scattering (reflection and transmission) of (Bloch)plane wave
states of the leads by the disordered segment of length $L\equiv N$ (with $a=1$).  For this purpose it is necessary to
perform a similarity transformation of (\ref{new5}) to the basis of the Bloch wave solutions 
$\varphi_n\sim e^{\pm ikn}$ for the leads.  The eigenvectors of $\widehat P^0$ are of the form 
$\vec u_{\pm}=\begin{pmatrix}e^{\pm ik}\\1\end{pmatrix}$ with eigenvalues $e^{\pm ik}$ obeying the equation 

\begin{equation}\label{new6}
E=2\cos k
\quad ,
\end{equation}
which defines the tight-binding energy band.  As usual we choose $k$ positive, $0\leq k\leq \pi$, so that e.g. $ e^{ikn}$ 
corresponds to a wave propagating from left to right on the lattice of (\ref{new2}).  The similarity transformation of
$\widehat P_n$ to the Bloch wave basis is defined by the matrix $\widehat U=(\vec u_+,\vec u_-)$ and leads to

\begin{equation}\label{new7}
\widehat Q_n=\widehat U^{-1}\widehat P_n\widehat U=\widehat Q^0_n+\widehat Q^1_n
\quad ,
\end{equation}
where
\begin{equation}\label{new8}
\widehat Q^0_n=
\begin{pmatrix}
(1-b)e^{ik} & -b e^{-ik}\\
b e^{ik} & (1+b) e^{-ik}
\end{pmatrix}
,\;
\widehat Q^1_n=ib_n
\begin{pmatrix}
e^{ik} & e^{-ik}\\
-e^{ik} & -e^{-ik}
\end{pmatrix}
\quad ,
\end{equation}
and
\begin{equation}\label{new9}
b=\frac{\eta}{2\sin k}\;,\;b_n=\frac{\varepsilon_n}{2\sin k}
\quad ,
\end{equation}

The transfer matrix of the disordered segment of length $N$ is the product of transfer matrices associated with the 
individual sites:
\begin{equation}\label{new10}
\widehat Q=\prod^N_{n=1}\widehat Q_n
\quad .
\end{equation}

We recall now the precise relationship between the transfer matrix elements $\left(\widehat Q\right)_{ij}\equiv Q_{ij}$ 
and the reflection and transmission amplitudes $r^{+-}$ and 
$t^{--}$, and $ r^{-+}$  and $ t^{++}$ for waves incident at the left and at tle right of the disordered system, 
respectively.  The reflection and transmission amplitudes define the scattering matrix $\widehat S$, which expresses
outgoing wave amplitudes at te left ($O$) and at the right ($O'$) of the disordered segment in terms of ingoing ones,
($I$) and ($I'$)\cite{20}:
\begin{equation}\label{new11}
\begin{pmatrix}
O \\O'
\end{pmatrix}=
\begin{pmatrix}
r^{-+} & t^{--}\\ t^{++} & r^{+-}
\end{pmatrix}
\begin{pmatrix}
I\\ I'
\end{pmatrix}
\quad .
\end{equation}
The transfer matrix $\widehat Q$, on the other hand, gives the Bloch wave amplitudes at the right end of the disordered 
section in terms of the amplitudes at the left end:
\begin{equation}\label{new12}
\begin{pmatrix}
O'\\ I'
\end{pmatrix}=
\widehat Q\begin{pmatrix}
I\\ O
\end{pmatrix}
\quad ,
\end{equation}
whose transformation to a form analogous to (\ref{new11})yields:

\begin{equation}\label{new13}
\begin{pmatrix}
O\\ O'
\end{pmatrix}=
\frac{1}{Q_{22}}
\begin{pmatrix}
-Q_{21} & 1 \\ \text{det}\;\widehat Q & Q_{12}
\end{pmatrix}
\begin{pmatrix}
I\\ I'
\end{pmatrix}
\quad ,
\end{equation}
which leads to the desired expressions of transmission- and reflection amplitudes in terms of the transfer matrix elements 
$Q_{ij}=\left(\widehat Q\right)_{ij}$

\begin{equation}\label{new14}
t^{--}=\frac{1}{Q_{22}},\; t^{++}=\left(\det\widehat Q\right) t^{--}\quad ,
\end{equation}

\begin{equation}\label{new15}
r^{+-}=\frac{Q_{12}}{Q_{22}},\; r^{-+}=-\frac{Q_{21}}{Q_{22}}
\quad .
\end{equation}
From (\ref{new10}) it follows that the determinant of $\widehat Q$ is the product of the determinants of the exact 
transfer matrices $\widehat Q_n$ associated with the individual sites $n$ of the disordered segment of $N$ sites, of
lengths $N\; a$.  Now, from (\ref{new7}-\ref{new8}) we find that $\det\widehat Q_n=1,\; n=1,2,\ldots N$ so that

\begin{equation}\label{new16}
\det\widehat Q=1
\quad .
\end{equation}

From (\ref{new14}) it then follows that 
\begin{equation}\label{new17}
t^{++}= t^{--}\equiv t=\frac{1}{Q_{22}}\quad ,
\end{equation}
for any realization of the disorder and for any strength of the imaginary potential.

As usual, we assume that the random site energies are identically distributed independent Gaussian variables with mean 
zero and correlation
\begin{equation}\label{new18}
\langle \varepsilon_n\varepsilon_m\rangle=\varepsilon^2_0 \delta_{m,n}
\quad .
\end{equation}
For weak disorder, we shall expand the matrix functional $\widehat Q$ to linear order in the random site energies.  We 
note that since the energies of neighboring sites are uncorrelated (Eq. (\ref{new18})) the second order terms in the
expansion of (\ref{new10}) may be omitted since they will not contribute in averages over the disorder at the order
$\varepsilon^2_0$.  Thus restricting to first order in the expansion of (\ref{new10}), we obtain\cite{20}

\begin{equation}\label{new19}
\widehat Q=\left(\widehat Q^0_n\right)^N
+\sum^N_{m=1}
\left(\widehat Q^0_n\right)^{m-1}
\widehat Q^1_m
\left(\widehat Q^0_n\right)^{N-m}
\equiv \widehat Q^0+\widehat Q^1
\quad .
\end{equation}

The transfer matrix product $=\left(\widehat Q^0_n\right)^N$ for the medium in the absence of disorder may be readily 
evaluated in closed form.  This will allow us to discuss analytically the reflection and transmission properties of a
perfectly amplifying or absorbing system, as done earlier by Datta\cite{12} using a slightly different procedure.  
We write

\begin{equation}\label{new20}
\left(\widehat Q^0_n\right)^N=
\widehat V \left(\widehat V^{-1}\widehat Q^0_n\widehat V\right)^N
\widehat V^{-1}, \;
\widehat V=(\vec v_{+},\vec v_{-})
\quad ,
\end{equation}
where $\widehat V$ is the diagonalizing matrix formed by the eigenvectors

\begin{equation}\label{new21}
\vec v_+=
\begin{pmatrix}
\frac{b e^{-ik}}{(1-b)e^{ik}-e^{iq}}\\ 1
\end{pmatrix}
,\;
\vec v_-=
\begin{pmatrix}
\frac{b e^{-ik}}{(1-b)e^{ik}-e^{-iq}}\\ 1
\end{pmatrix}
\quad ,
\end{equation}
(with $\text{det}\;\widehat V=\frac{2i\ e^{-ik}}{b^2}\sin q$) of $\widehat Q^0_n$ corresponding to eigenvalues $e^{iq}$ 
and $e^{-iq}$, respectively.  These eigenvalues are defined in terms complex functions $\cos q$ and $\sin q$ by

\begin{eqnarray}\label{new22}
e^{\pm iq}&=&\cos q\pm i\sin q,\nonumber\\
\cos q &=& \cos k-ib\sin k,\; \sin q=\sqrt{1-\cos^2 q}
\quad .
\end{eqnarray}
The explicit expression of $\left(\widehat Q^0_n\right)^m$ obtained from (\ref{new20}-\ref{new22}) is

\begin{equation}\label{new23}
\left(\widehat Q^0_n\right)^m=
\begin{pmatrix}
A_m & B_m\\ C_m & D_m
\end{pmatrix}
\quad ,
\end{equation}
where (with $e^{\pm ik}\equiv d_\pm$)

\begin{equation}\label{new24}
A_m=
\frac{1}{\sin q}[(1-b)d_+ \sin q m-\sin q(m-1)]
\quad ,
\end{equation}

\begin{equation}\label{new25}
D_m=
-\frac{1}{\sin q}[(1-b)d_+ \sin q\; m-\sin q(m+1)]
\quad ,
\end{equation}

\begin{equation}\label{new26}
B_m=
-b\;d_-\frac{\sin q\; m}{\sin q}
\quad ,
\end{equation}

\begin{equation}\label{new27}
C_m=
b\;d_+\frac{\sin q\; m}{\sin q}
\quad .
\end{equation}
Note that, as expected, the transfer matrix of the non-disordered system with absorption or gain defined by 
(\ref{new24}-\ref{new27}) obeys

\begin{equation}\label{new28}
A_N D_N -B_N C_N=\cos^2 qN +\sin^2 qN=1
\quad ,
\end{equation}

and from (\ref{new15}) and (\ref{new26}-\ref{new27}) it follows that in this case the reflection coefficients are 
equal

\begin{equation}\label{new29}
|r^2|\equiv |r^{+-}|^2=|r^{-+}|^2=
\frac{|B_N|^2}{|D_N|^2},\;
\varepsilon_m=0,\;
m=1,2,\ldots N
\quad .
\end{equation}

Finally, by inserting (\ref{new8}) and (\ref{new23}) in (\ref{new19}), we obtain the explicit form of the transfer matrix 
of the disordered section to first order:

\begin{equation}\label{new30}
\widehat Q=
\begin{pmatrix}
A_N & B_N \\ C_N & D_N
\end{pmatrix}+
\begin{pmatrix}
Q^1_{11} & Q^1_{12}\\ Q^1_{21} & Q^1_{22}
\end{pmatrix}
\quad ,
\end{equation}
where
\begin{equation}\label{new31}
Q^1_{11}=
i\sum^N_{m=1} b_m
(A_{m-1}-B_{m-1})(d_+ A_{N-m}+d_-C_{N-m})
\quad ,
\end{equation}

\begin{equation}\label{new32}
Q^1_{22}=
i\sum^N_{m=1} b_m
(C_{m-1}-D_{m-1})(d_+ B_{N-m}+d_-D_{N-m})
\quad ,
\end{equation}

\begin{equation}\label{new33}
Q^1_{12}=
i\sum^N_{m=1} b_m
(A_{m-1}-B_{m-1})(d_+ B_{N-m}+d_-D_{N-m})
\quad ,
\end{equation}

\begin{equation}\label{new34}
Q^1_{21}=
i\sum^N_{m=1} b_m
(C_{m-1}-D_{m-1})(d_+ A_{N-m}+d_-C_{N-m})
\quad ,
\end{equation}
The lowest order effect of the disorder in the mean transmission and reflection coefficients is obtained by expanding 
(\ref{new15}) and (\ref{new17}) to second order in the disorder, using (\ref{new30}) and noting that the first order
averages vanish.  However in the case of the reflection coefficients, which are asymptotically independent of length in
the absence of disorder\cite{12,13}, it is apt to focus on the simpler averages $\langle\ln|r^{\pm\mp}|^2\rangle$.  On the
other hand, the study of the mean logarithm of the transmission coefficient is of special interest since it is related
asymptotically to the localization length

\begin{equation}\label{new35}
\frac{1}{\xi_\pm}=-\lim\frac{\langle\ln |t^{\pm\pm}|^2\rangle}{2N}
\quad ,
\end{equation}
which is a self-averaging quantity in the absence of absorption or gain\cite{21}.  From (\ref{new15}), (\ref{new17}) and 
(\ref{new30}) we obtain successively for the quantities of interest, to second order in the disorder,

\begin{equation}\label{new36}
\langle\ln |t|^2\rangle=
(-\ln D_N+c.c.)+
\frac{1}{2}
\left( \frac{\langle(Q^1_{22})^2\rangle}{D^2_N}+ c.c.\right)
\quad ,
\end{equation}

\begin{equation}\label{new37}
\langle\ln |r^{+-}|^2\rangle=
\left(\ln \frac{B_N}{D_N}+ c.c.\right)-\frac{1}{2}
\left[\left(\langle\left(\frac{Q^1_{12}}{B_N}\right)^2\rangle
-\langle\left(\frac{Q^1_{22}}{D_N}\right)^2\rangle\right)+c.c.\right]
\quad ,
\end{equation}

\begin{equation}\label{new38}
\langle\ln |r^{-+}|^2\rangle=
\left(\ln \frac{C_N}{D_N}+ c.c.\right)-\frac{1}{2}
\left[\left(\langle\left(\frac{Q^1_{21}}{C_N}\right)^2\rangle
-\langle\left(\frac{Q^1_{22}}{D_N}\right)^2\rangle\right)+c.c.\right]
\quad .
\end{equation}

We close this section by demonstrating the equivalence of our results for the transmission- and reflection coefficients 
for the perfect absorbing or amplifying system ($\varepsilon_m=0$) and the corresponding results obtained earlier by
Datta\cite{12} at the band center.  For this purpose we identify the parameters $e^{iq}$ and 
$e^{-iq}$ defined above respectively with the quantities $i\; e^{-\xi}$ and $-i\; e^\xi$ involving the parameter $\xi$ 
introduced by Datta via the substitution $\sinh\xi=\frac{\eta}{2}$.  By transforming Eq. (5) of\cite{12} for $|t|^2$ for
even $N$ in terms of the variable $q$ we get

\begin{equation}\label{new39}
|t|^2=
\frac{\sin^2 q}{(-i\sin Nq+\sin q\cos Nq)^2}
\quad ,
\end{equation}
which coincides with the expression obtained by substituting (\ref{new25}) for $k=\pi/2$ and 
$m=N$ in the definition (\ref{new17}) of $|t^{--}|$.  Similarly, the transformation of Eq. (6) of\cite{12} for odd $N$ 
yields again (\ref{new39}) obtained from (\ref{new17}) and (\ref{new25}) above.  The equations (7) and (8) of
Datta\cite{12} for the reflection coefficient $|r|^2$ for even and odd $N$, respectively, reduce similarly to the
corresponding expression obtained from (\ref{new25}-\ref{new26}) and (\ref{new29}). Clearly, the advantage of the present
treatment is that it condenses distinct expressions for even and odd $N$ in Datta's analysis into a single one for
any one of the amplitudes coefficients in (\ref{new15}), (\ref{new17}).  This is clearly useful, particularly for handling
the more cumbersome general expressions for the effect of a weak disorder.

\section{DETAILED RESULTS FOR $E=0$}

For simplicity and as in most previous work for the tight-binding model\cite{11,12,13}, we restrict the analytical 
calculations and results in this section to the band center, $E=0(k=\pi/2)$.  At the band center the pure system transfer
matrix elements (\ref{new24}-\ref{new27}) take the simple forms

\begin{equation}\label{new40}
A_m=u_+ e^{iqm}-u_-e^{-iqm}
\quad ,
\end{equation}

\begin{equation}\label{new41}
D_m=-u_- e^{iqm}+u_+e^{-iqm}
\quad ,
\end{equation}

\begin{equation}\label{new42}
B_m=C_m=v(e^{iqm}- e^{-iqm}),\; m=1,2,\ldots,N
\quad ,
\end{equation}
where

\begin{equation}\label{new43}
u_{\pm}=
\frac{1}{2}
\left(
\frac{1}{\sqrt{1+b^2}}\pm 1
\right),\;
v=\frac{b}{2\sqrt{1+b^2}}
\quad ,
\end{equation}

\begin{equation}\label{new44}
e^{\pm iq}=(\pm i)(\sqrt{1+b^2}\mp b)
\quad .
\end{equation}

For the pure tight-binding system with absorption or gain ($\varepsilon_m=0,m=1,2\ldots N$) the transmittances and 
reflectances, for both directions of incidence, are given exactly for any band energy $E$ and for any length $L=N$ by
substituting the closed expression (\ref{new24}-\ref{new27}) for the transfer matrix elements into the definition
(\ref{new15}) and (\ref{new17}).  Exact results for $|t|^2$ and $|r|^2$ for the perfect system with absorption or gain, for
$E=0$, have been discussed by Datta\cite{12} and more extensive numerical results which include the additional effect of a
weak disorder on the averaged logarithmic transmittance have been presented by Jiang and Soukoulis\cite{13}.  Special
attention has been paid in\cite{12,13} to the domain of intermediate lengths (in particular the critical length $L_c$)
where the transmittance of an amplifying system changes from an initial growth at short lengths to an exponential decay at
long lengths.

In the following we discuss detailed results for transmittance and reflectance in the framework of the general analytic 
treatment for weak disorder in Sect. II.  We shall consider successively the short- and the long lengths domains defined
below.  Our consistent treatment of the effect of a weak disorder in the framework of an exact analysis of absorption or
gain at zeroth order leads to the identification of the important effects induced by absorption or amplification in the
statistics of wave transport.

\subsection{Short lengths}

For a fixed magnitude of the absorption/amplification parameter $b$ the short lengths domain is defined by

\begin{equation}\label{new45}
N|b|<< 1
\quad ,
\end{equation}
or, equivalently $L<<l_0$ where $l_0=1/b$ for $b>0$ is the absorption length (in units of $a$) and $l_0=-1/b$ for $b<0$ is 
the amplification length.  We wish to obtain the logarithmic transmittance in the limit (\ref{new45}), which determines
the short-length localization length.  For this purpose we use the following approximations of (\ref{new40}-\ref{new42})
valid to lowest order, for small $|b|$ and small $m|b|$,

\begin{equation}\label{new46}
A_m=i^m(1-mb)
\quad ,
\end{equation}

\begin{equation}\label{new47}
D_m=(-i)^m(1+mb)
\quad ,
\end{equation}

\begin{equation}\label{new48}
B_m=C_m=\frac{b}{2}(i^m+(-i)^m)
\quad .
\end{equation}
Note that these expressions would not be sufficient for discussing the reflection coefficients whose explicit forms differ 
for even and odd $N$ and require inclusion of higher orders in $ma\equiv m$ for $B_m$ and $C_m$.  For brievety's sake
we omit discussing the short lengths reflection coefficients in more detail.  We substitute (\ref{new46}-\ref{new48}) in
the expression (\ref{new32}) for $E=0$, which we then insert in (\ref{new36}).  After averaging over the disorder, using
(\ref{new18}), and performing the remaining geometric sums over sites, we obtain the following final results:

\begin{equation}\label{new49}
\langle\ln |t|^2\rangle=
-2bN
-\frac{\varepsilon^2_0}{4}
(1-4b)N+O(b^2N^2)
\quad ,
\end{equation}

The short-length localization length obtained from (\ref{new35}) and (\ref{new49}) namely,

\begin{equation}\label{new50}
\frac{1}{\xi}=b+\frac{\varepsilon^2_0}{8}
(1-4b)
\quad ,
\end{equation}
yields
\begin{equation}\label{new51}
\frac{1}{\xi}=\frac{1}{l_0}+\frac{1}{\xi_0}
-\frac{4}{\xi_0 l_0}
\quad ,
\end{equation}
for absorption, and
\begin{equation}\label{new52}
\frac{1}{\xi}=-\frac{1}{l_0}+\frac{1}{\xi_0}
+\frac{4}{\xi_0 l_0}
\quad ,
\end{equation}
for amplification, where

\begin{equation}\label{new53}
\xi_0=\frac{8}{\varepsilon^2_0}
\quad .
\end{equation}
Note that (\ref{new53}) is the exact perturbation expression (for $E=0$) of the localization length for weak disorder, for 
$b=0$.  Indeed it coincides with the well-known exact result, 
$\xi_0=96W^{-2}\sin^2k$, obtained by Thouless\cite{22}, if the variance $W^2/12$ of the rectangular distribution of width 
$W$ of site energies in\cite{22} is replaced by the gaussian mean square $\varepsilon^2_0$.  The first two terms in
(\ref{new51}-\ref{new52}) agree with the form of the short-lengths localization lengths derived previously from invariant
imbedding \cite{4,5,10}.

Finally, it is useful to clarify the general meaning of short- and long-lengths localization lengths in the framework of 
our weak disorder analysis.  We recall that the perturbation treatment of disorder in Sect. II is valid for 

\begin{equation}\label{new54}
N\varepsilon^2_0<<1
\quad .
\end{equation}
This implies, in particular, that the localization length (\ref{new53}) is valid for values $\varepsilon^2_0\rightarrow 0$ 
such that the limit (\ref{new54}) embraces asymptotically large $N$ for which the localization length is defined in
(\ref{new35}).  Similarly, the short-lengths localization length (\ref{new50}) in the presence of absorption or gain is a
true localization length only if it corresponds to the limit of asymptotically large $N$ in (\ref{new35}).  Thus if
$|b|<\varepsilon^2_0$ this limit is obtained for $|b|\rightarrow 0$ (since $N|b|<N\varepsilon^2_0<<1$) while if
$|b|>\varepsilon^2_0$ it is obtained by letting $\varepsilon^2_0\rightarrow 0$.  Now, for $|b|<\varepsilon^2_0$
(\ref{new54})automatically implies (\ref{new45}), in which case the short-lengths expressions (\ref{new51}-\ref{new52})
give the true localization lengths.  On the other hand, for $|b|>\varepsilon^2_0$ two possibilities exist for the
localization lengths:

\begin{itemize}
\item if for asymptotic lengths obeying (\ref{new54})($\varepsilon^2_0\rightarrow 0$) one also has 
$N|b|<<1$, then the localization lengths are clearly given by the "short-lengths" expressions (\ref{new51}-\ref{new52}).  
This happens for values $|b|\geq\varepsilon^2_0$ sufficiently close to $\varepsilon^2_0$.

\item if for asymptotic lengths (\ref{new54}) the long lengths condition
\end{itemize}

\begin{equation}\label{new55}
N|b|>>1
\quad .
\end{equation}
is fulfilled, then the localization lengths are given by the equations (\ref{eq:mynum1},\ref{eq:mynum3}) and their
limiting  forms (\ref{eq:mynum5}, \ref{eq:mynum7}) and 
(\ref {eq:mynum9},\ref{eq:mynum11}) in subsection B below.  This situation exists for $|b|$-values sufficiently larger
than 
$\varepsilon^2_0$.

\subsection{Long lengths}
The transfer matrix elements (\ref{new40}-\ref{new42}) of a perfect system depend on the imaginary exponentials 
$e^{\pm iqN}$, which for $|b|<<1$ are given by 

\begin{equation}\label{new56}
e^{\pm iqN}=(\pm i)^N e^{\mp N(b+\frac{b^3}{3}+\ldots)}
\quad ,
\end{equation}
where $e^{iqN}$ is exponentially growing for $b<0$ (amplification) and $e^{-iqN}$ is growing for $b>0$ (absorption) in the 
long lengths regime, $N>>|b|^{-1}$ (\ref{new55}).  We first discuss the detailed form of the logarithmic transmission
coefficient, 
$\ln |t|^2=\ln t+ c.c.$, and of the reflection coefficient $|r^{\pm\mp}|^2$ given by (\ref{new17}),(\ref{new15}) and 
(\ref{new30}-\ref{new34}) in terms of the transfer matrix elements $A_N, B_N, C_N, D_N$ in (\ref{new40}-\ref{new42}). 
Retaining only the leading exponential terms at long lengths for absorption and amplification, respectively, we obtain
successively

\begin{equation}\label{new57}
\ln |t|^2=-2\left|\left(b+\frac{b^3}{3}\right)\right|N+O (N^{-1})
\quad ,
\end{equation}
for both signs of $b$, and

\begin{equation}\label{new58}
|r^{+-}|^2=|r^{-+}|^2\simeq\frac{b^2}{4}, b>0
\quad ,
\end{equation}

\begin{equation}\label{new59}
|r^{+-}|^2=|r^{-+}|^2\simeq\frac{4}{b^2},  b<0
\quad .
\end{equation}
The main feature of these results is that $\ln|t|^2$ is decreasing at large $L$ for amplification as well as for 
absorption in agreement with previous studies\cite{6,9,11,12,13}.

Next we consider the effect of weak disorder at $E=0$ in the mean logarithmic transport coefficients 
(\ref{new36}-\ref{new38}) involving zeroth- and first order transfer matrix elements defined in (\ref{new40}-\ref{new42})
and (\ref{new31}-\ref{new34}).  Using (\ref{new18}), the averages of the various quadratic forms in first order transfer
matrix elements in (\ref{new36}-\ref{new38}) reduce to simple sums over lattice sites $m$ of products of two terms of the
form $M_{m-1}-N_{m-1}$ corresponding to the site $m-1$ multiplied by a product of two terms of the form $
P_{N-m}-R_{N-m}$ corresponding to the site $N-m$ (with $M,N,P,R$ representing elements, distinct or not, of the set of
transfer matrix elements, $A,B,C,D$, of the pure system).  Using (\ref{new40}-\ref{new42}), we approximate the $m$th term
in a given sum by the contribution which is independent of $m$, which yields the leading effect proportional to $N$ for
any of the sums involved (the contributions ignored in this approximation are readily shown to be of relative order
$\frac{1}{N}$).  For the averages of products of first order transfer matrix elements entering in (\ref{new36}-\ref{new38})
we thus obtain the following results valid at $E=0$, for any sign of $b$:

\begin{equation}\label{new60}
\langle\left(Q_{22}^{1}\right)^2\rangle=
\frac{N\varepsilon^2_0}{4}
\left[(u_-+v)^2e^{iq(N-1)}+ [(u_++v)^2e^{-iq(N-1)}\right]^2
\quad ,
\end{equation}

\begin{equation}\label{new61}
\langle\left(Q_{12}^{1}\right)^2\rangle=
\frac{N\varepsilon^2_0}{4}
\left[(u_+-v) (u_-+v)
e^{iq(N-1)}+ [(u_++v) (u_--v)
e^{-iq(N-1)}\right]^2
\quad ,
\end{equation}

\begin{equation}\label{new62}
\langle\left(Q_{21}^{1}\right)^2\rangle=
\langle\left(Q_{12}^{1}\right)^2\rangle
\quad .
\end{equation}

From (\ref{new62}) and (\ref{new37}-\ref{new38}), it follows that

\begin{equation}\label{new63}
\langle \ln|r^{+-}|^2 \rangle=
\langle \ln|r^{-+}|^2 \rangle\equiv
\langle \ln|r|^2 \rangle
\quad ,
\end{equation}
for both signs of $b$.

Next we insert (\ref{new60}-\ref{new62}), together with (\ref{new41}-\ref{new42}) in (\ref{new35}-\ref{new38}) and 
simplify the resulting expressions by retaining in each one of them only the leading exponential terms for $N|b|>>1$,
successively for $b>0$ and $b<0$.  In this way we obtain the following exact expressions valid for arbitrary $|b|$ larger
than $\varepsilon^2_0$ and such that $N|b|>>1$:

\begin{equation}\tag{64.a}\label{eq:mynum1}
\frac{1}{\xi}=-\frac{1}{2N}\ln\vert e^{-iqN}\vert^2-\frac{\varepsilon^2_0}{8}
\frac{(u_++v)^4}{u^2_+} e^{2iq}
\quad ,
\end{equation}

\begin{equation}\tag{65.a}\label{eq:mynum2}
\langle\ln\vert r\vert^2\rangle=\ln \left(\frac{v}{u_+}\right)^2
-\frac{\varepsilon^2_0 N}{4}(u_++v)^2
\left[\frac{(u_--v)^2}{v^2}+\frac{(u_++v)^2}{u^2_+}\right] e^{2iq},\; b>0
\quad ;
\end{equation}

\begin{equation}\tag{64.b}\label{eq:mynum3}
\frac{1}{\xi}=-\frac{1}{2N}\ln\vert e^{iqN}\vert^2-\frac{\varepsilon^2_0}{8}
\frac{(u_-+v)^4}{u^2_-} e^{-2iq}
\quad ,
\end{equation}

\begin{equation}\tag{65.b}\label{eq:mynum4}
\langle\ln\vert r\vert^2\rangle=\ln \left(\frac{v}{u_-}\right)^2
-\frac{\varepsilon^2_0 N}{4}(u_-+v)^2
\left[\frac{(u_+-v)^2}{v^2}+\frac{(u_-+v)^2}{u^2_-}\right] e^{-2iq},\; b<0
\quad .
\end{equation}

For weak absorption/amplification $\vert b\vert <<1$ (with, however, $N\vert b\vert >>1$) we expand 
(\ref{eq:mynum1},\ref{eq:mynum2}) and (\ref{eq:mynum3},\ref{eq:mynum4})in powers of $b$, using (\ref{new43}-\ref{new44})
and (\ref{new56}).  To order $b^2$ in the effects of the disorder we obtain finally

\begin{equation}\tag{66.a}\label{eq:mynum5}
\frac{1}{\xi}=b+\frac{b^3}{3}+\frac{\varepsilon^2_0}{8}(1-b^2)
\quad ,
\end{equation}

\begin{equation}\tag{67.a}\label{eq:mynum6}
\langle\ln\vert r\vert^2\rangle=
\ln\left(\frac{b^2}{4}\right)-\frac{b^2}{2}+\frac{\varepsilon^2_0 b^2}{2}N,\; b>0
\quad ;
\end{equation}

\begin{equation}\tag{66.b}\label{eq:mynum7}
\frac{1}{\xi}=-\left(b+\frac{b^3}{3}\right)+\frac{\varepsilon^2_0}{8}(1-b^2)
\quad ,
\end{equation}

\begin{equation}\tag{67.b}\label{eq:mynum8}
\langle\ln\vert r\vert^2\rangle=
\ln\left(\frac{4}{b^2}\right)+\frac{b^2}{2}, \; b<0
\quad .
\end{equation}

On the other hand, for strong absorption/amplification, $\vert b\vert >>1$, we obtain successively from 
(\ref{eq:mynum1},\ref{eq:mynum2}) and (\ref{eq:mynum3},\ref{eq:mynum4}), to the orders indicated,

\begin{equation}\tag{68.a}\label{eq:mynum9}
\frac{1}{\xi}=b+\frac{b^3}{3}+\frac{\varepsilon^2_0}{8b^2}
\quad ,
\end{equation}

\begin{equation}\tag{69.a}\label{eq:mynum10}
\langle\ln\vert r\vert^2\rangle=-\frac{2}{b}
\left(1+\frac{1}{6b^2}\right)
+\frac{\varepsilon^2_0 N}{2b ^2}
\left(1-\frac{1}{\vert b\vert}\right)
,\; b>0
\quad ;
\end{equation}

\begin{equation}\tag{68.b}\label{eq:mynum11}
\frac{1}{\xi}=-\left(b+\frac{b^3}{3}\right)+\frac{\varepsilon^2_0}{8b^2}
\quad ;
\end{equation}

\begin{equation}\tag{69.b}\label{eq:mynum12}
\langle\ln\vert r\vert^2\rangle=\frac{2}{b}
\left(1-\frac{1}{3b^2}\right)
+\frac{\varepsilon^2_0 N}{2b ^2}
\left(1+\frac{1}{\vert b\vert}\right)
,\; b<0
\quad ,
\end{equation}
using  expansions of (\ref{new43}-\ref{new44}) in powers if $\frac{1}{\vert b\vert}$.

Our detailed results (\ref{eq:mynum5},\ref{eq:mynum7}) and 
(\ref{eq:mynum9},\ref{eq:mynum11}) for localization lengths and (\ref{eq:mynum6},\ref{eq:mynum8}) and 
(\ref{eq:mynum10},\ref{eq:mynum12}) for logarithmic reflection coefficients display remarkable new features related to the
effects induced by absorption/amplification in statistical averages over the disorder. We recall that our results are
valid at asymptotic lengths for values $\vert b\vert>\varepsilon^2_0$ but not too close to $\varepsilon^2_0$ (see the
discussions in III.A above). In the absence of the induced statistical  effects the results for inverse localization
lengths coincide with the previously known results (1), (3)\cite{6,9}.  On the other hand, in the absence of disorder, the
results for the localization length (transmission coefficient) and for the reflection coefficient coincide with the exact
results obtained by Datta\cite{12}.  In particular, in the absence of disorder the reflection coefficient is
asymptotically constant for $N\rightarrow\infty$\cite{12,13}.

Concerning now the statistical effects induced by absorption/amplification in the inverse localization lengths, our 
results (\ref{eq:mynum5},\ref{eq:mynum7}) and (\ref{eq:mynum9},\ref{eq:mynum11}) lead to the following conclusions:

\begin{enumerate}
\item 
the effects are identical for absorption and for amplification, for weak- as well as for strong absorption/amplification.

\item
the statistical effect induced by absorption/amplification increases the localization length for weak 
absorption/amplification while reducing it for strong absorption/amplification.
\item
localization by disorder is destroyed in the presence of sufficiently strong absorption/amplification.
\end{enumerate}

On the other hand, our results in (\ref{eq:mynum6},\ref{eq:mynum8}) and 
(\ref{eq:mynum10},\ref{eq:mynum12}) for the statistical effects induced by absorption/amplification in the reflection 
coefficient reveal that:

\begin{enumerate}
\item
a weak absorption induces a weak asymptotic statistical growth of $\langle\vert r\vert ^2\rangle$ while a corresponding 
weak amplification leads to no statistical effect.
\item
for large absorption/amplification parameters $\vert b\vert$, absorption and amplification induce identical weak 
statistical growth terms (to leading order in $\vert b\vert ^{-2}$) in 
$\langle\vert r\vert ^2\rangle$.
\end{enumerate}

\section{CONCLUDING REMARKS}

The main results of this paper are summarized in the analytical expressions (\ref{new51}-\ref{new52}) (\ref{eq:mynum5},
\ref{eq:mynum7}-\ref{eq:mynum10},\ref{eq:mynum12}) for inverse localization lengths and logarithmic reflection
coefficients in short- and long random tight-binding systems with absorption or gain.  These results are discussed in
detail in the main text.  Our analysis in Section II is valid for $N\varepsilon^2_0<<1$ which characterizes the weak
localization regime identified more generally by the limit $L<<\xi_0$.  It would be interesting in future work to study
the effect of absorption or amplification in the strong localization (or localized-) regime $L>>\xi_0$, for weak
disorder.  This would allow to study the additional effects associated with anomalies in $\xi_0$ existing at
special energies, in particular at the band center\cite{23}.  The study of transmission and reflection in the localized
regime requires a more involved treatment of the disorder, respecting, in particular, the asymptotic unitarity limit of the
reflection coefficient in the absence of absorption or gain.  A simple analytic treatment of statistical properties of the
transmittance in the localized regime in the absence of absorption has been discussed recently in\cite{24}.

We close with brief remarks on the respective roles of different symmetries of the transfer matrix (or of the lack thereof)
 for the disordered tight-binding system with absorption or gain studied above.  In Sect. II a central role is played by
the transfer matrix $\widehat Q_n$ of an elementary disordered segment enclosing just one site $n$.  $\widehat Q_n$ obeys
the property

\begin{equation}\tag{70}\label{new70}
\det \widehat Q_n=1
\quad ,
\end{equation}
which leads to the relations

\begin{equation}\tag{71}\label{new71}
\vert t_n^{++}\vert^2=\vert t_n^{--}\vert^2,\; \vert r_n^{+-}\vert^2=\vert r_n^{-+}\vert^2
\quad ,
\end{equation}
for the reflection- and transmission coefficients for waves incident from the left and from the right, respectively.  
As shown in Sect. II, (\ref{new70}) implies that $\det\widehat Q=1$ which in turn leads to the identity of the transmission
coefficients $\vert t^{++}\vert^2$ and 
$\vert t^{--}\vert^2$ (Eq. (\ref{new17})) for a system of $N$ sites.  Now (\ref{new71}) may be viewed simply as reflecting 
the symmetry of the piecewise defined solutions of the Schr\"{o}dinger equation for plane waves incident from the right and
from the left, respectively, in a single-site random segment.  This finally shows that the transfer matrix $\widehat Q$
embodies the basic left-right symmetry of Eq.(\ref{new2}), via Eq.(\ref{new16}), which leads to the properties 
$\langle\vert t^{++}\vert^2\rangle=\langle\vert t^{--}\vert^2\rangle\equiv\langle\vert t\vert^2\rangle $ 
$\langle\vert r^{+-}\vert^2\rangle =\langle\vert r^{-+}\vert^2\rangle \equiv\langle\vert r\vert^2\rangle$ for the
observable transmission and reflection coefficients.

In the absence of absorption or gain the disordered system (\ref{new2}) possesses a further well-known symmetry, namely 
time-reversal symmetry.  This symmetry implies that the $2\times 2$ transfer matrix $\widehat X$ satisfies the condition

\begin{equation}\tag{72}\label{new72}
\widehat X^*=\sigma\widehat X\sigma,\; \sigma=
\begin{pmatrix}
0&1\\1&0
\end{pmatrix}
\quad .
\end{equation}
This symmetry is broken when absorption or gain is present as follows e.g. from the transfer matrix (\ref{new23}) for the 
pure system.  Indeed, for $E=0$, (\ref{new72}) would require that 

\begin{equation}\tag{73}\label{new73}
B_N^*=C_N\;\text{and}\; A_N^*=D_N
\quad ,
\end{equation}
which is not the case for the elements (\ref{new40}-\ref{new42}).  Now it is readily seen that the lack of time-reversal 
symmetry as shown by the violation of (\ref{new73}) is related to a physical fact, namely the absence of current
conservation, which means that 
$\vert r\vert^2+\vert t\vert^2\neq 1$.  Indeed, from (\ref{new17}) and (\ref{new29}) we  have in the present case

\begin{equation}\tag{74}\label{new74}
\vert r\vert^2+\vert t\vert^2=
\frac{1+\vert B_N\vert^2}{\vert D_N\vert^2}\neq 1
\quad ,
\end{equation}
as seen from (\ref{new28}), since $A_N\neq D_N^*$ and $C_N\neq B_N^*$.  Note also a further related consequence of the 
lack of time-reversal symmetry of the perfectly absorbing or amplifying systems:  this is the violation of the duality
relation for the scattering-matrix derived by Paaschens {\it et al.}\cite{9}.  Violation of the duality relation
of\cite{9} for the $S$-matrix is easily demonstrated by substituting the transfer matrix elements
(\ref{new41}-\ref{new43}) in Eqs. (\ref{new14}-\ref{new15}) for the reflection and transmission amplitude coefficients.


\begin{thebibliography}{unsrt}
\bibitem{16} The random potential is assumed to be inhomogeneous in the $x$-direction only and includes a fixed imaginary
part describing absorption ($\eta>0$) or gain ($\eta<0$).
\bibitem{19} In the appendix of \cite{10} a new simple derivation of the invariant imbedding equations from the Helmholtz 
equation was presented.  This derivation depends on the identification of $dy (L,L)/dL$ (where
$y(x,L)=\frac{\partial}{\partial x}\ln E(x,L)$, with $E(x,L)$ the electric wave field)  with 
$\left(\frac{\partial y(x,L)}{\partial x}\right)_{x=L}$ at the edge $x=L$ of the layer extending between $x=0$ and $x=L$.
In
\cite{10} this identification was regarded as "a basic assumption of the invariant imbedding procedure".  Actually, it
turns out that this is not an assumption at all, since it may be shown, using the exact identity 
$\frac{\partial E(x,L)}{\partial L}=a(L)E(x,L)$, $a(L)=ik_0 \left[1+\frac{1}{2}\eta (L)E(L,L)\right]$, derived in\cite{4},
that
$\left(\frac{\partial y(x,L)}{\partial L}\right)_{x=L}=0$.   In conclusion the derivation of the invariant imbedding
equations in\cite{10} is exact.
\end{thebibliography}
\end{document}